\documentclass[12pt]{article}
\usepackage{latexsym}
\usepackage{amsmath}
\usepackage{hyperref}
\usepackage{amssymb}
\usepackage{cite}

\DeclareMathAlphabet{\mathpzc}{OT1}{pzc}{m}{it}
\newcommand{\alter}[1]{{\mathpzc{#1}}}

\newcommand{\be}{\begin{equation}}
\newcommand{\ee}{\end{equation}}
\newcommand{\bea}{\begin{eqnarray}}
\newcommand{\eea}{\end{eqnarray}}
\newcommand{\ba}{\begin{array}}
\newcommand{\ea}{\end{array}}
\newcommand{\ds}{\displaystyle}

\renewcommand{\L}{\Lambda}

\newcommand{\diff}{{\tt d}}
\newcommand{\der}{\partial}

\newcommand{\x}{\psi}  
\newcommand{\dilaton}{\phi} 
\newcommand{\scal}{\varphi} 
\newcommand{\scalC}{X}      
\newcommand{\Mf}{{\cal M}}  
\newcommand{\Nf}{{\cal N}}  
\newcommand{\h}{h}          
\newcommand{\znam}{R}          
\newcommand{\chisl}{S}          

\newcommand{\dtensor}{d}   

\newcommand{\fl}[1]{{{\hat{#1}}}} 
\newcommand{\eig}{\lambda}  
\newcommand{\tetrad}{e}  

\newcommand{\hepthOld}[1]{\href{http://lanl.arxiv.org/abs/hep-th/#1}{arXiv:hep-th/#1}}
\newcommand{\hepthNew}[1]{\href{http://lanl.arxiv.org/abs/#1}{arXiv:#1 [hep-th]}}

\begin{document}

\begin{titlepage}

    \thispagestyle{empty}
    \begin{flushright}
        \hfill{CERN-PH-TH/2010-138}
    \end{flushright}

\vspace{1em}
\begin{center}
{\LARGE{\textbf{
Attractors and first order formalism\\[0.5em] in five dimensions revisited
}}}\\

\vspace{2em}

{{\textbf{S. Bellucci${}^1$, S. Ferrara${}^{1,2,3}$, A. Shcherbakov${}^{4,1}$, A. Yeranyan${}^{1,5}$}}}\\

\vspace{0.5em}
{${}^1$ \it INFN - Laboratori Nazionali di Frascati, \\
Via Enrico Fermi 40,00044 Frascati, Italy\\
\texttt{bellucci, ashcherb, ayeran@lnf.infn.it}}

\vspace{0.5em}
{${}^2$ \it Physics Department, Theory Unit, CERN, \\
CH 1211, Geneva 23, Switzerland\\
\texttt{sergio.ferrara@cern.ch}}

\vspace{0.5em}
{${}^3$ \it Department of Physics and Astronomy,\\
 University of California, Los Angeles, CA USA}

\vspace{0.5em}
{${}^4$ \it Museo Storico della Fisica e\\
        Centro Studi e Ricerche ``Enrico Fermi"\\
        Via Panisperna 89A, 00184 Roma, Italy}

\vspace{0.5em}
{${}^5$ \it Department of Physics, Yerevan State University,\\
Alex Manoogian St., 1, Yerevan, 0025, Armenia}

\end{center}
\vspace{1em}
\begin{abstract}
\vspace{0.5em} The attractor mechanism in five dimensional
Einstein-Maxwell Chern-Simons theory is studied. The expression of
the five dimensional rotating black object potential depending on
Taub-NUT, electric and magnetic charges as well as on all the scalar
and gauge fields, is investigated. The first order formalism in~$d=5$
is constructed and analyzed. We derive a general
expression defining the fake superpotential which is valid for all
charge configurations. An
explicit expression for the fake superpotential is constructed, for
all very special geometries, in the case of vanishing Taub-NUT
charge. We carry out an analogous construction in the very
special geometries corresponding to~$t^3$ and~$stu$ models, for the most general charge
configurations. The attractor flows and horizon values of all fields
are given.
\end{abstract}
\end{titlepage}
\date{}

\section*{Introduction}

The study of the low energy approximation is usually worthwhile, in order to clarify some peculiar properties of the full intricate theory, since in many cases it is able to grasp the main features of the theory. This happens, for example, in the case of superstring theory compactified on~$CY_3$ manifold, which yields a four dimensional supergravity theory. It is also the case for the same type of compactification of~$M$-theory, where one ends up with a five dimensional supergravity theory.

Supergravity solutions in five space-time dimensions often reveal
interesting connections with their four dimensional counterparts. A vast literature is devoted to this
issue~\cite{Cvetic:1996xz,Cvetic:1998xh,Larsen:2006xm,Castro:2008ne,%
Berkooz:2008rj,Loran:2008mm,Kraus:2005gh,Bena:2005ni,Gutowski:2008tm,%
LopesCardoso:2007ky,Cardoso:2007rg,Goldstein:2007km,Gao:2008hw,Ferrara:2006xx,%
Gaiotto:2005xt,Gaiotto:2005gf,Behrndt:2005he,Ceresole:2007rq,Hornlund:2010tr}.

Investigating black hole solutions in five dimensional gravity theory is interesting from different points of view. First of all,  five dimensional supergravity can be thought of as the eleven dimensional one compactified on a six torus~$T^6$ or a six dimensional Calabi-Yau space. Secondly, the five dimensional theory is the theory of highest dimensionality in which supersymmetric black holes might reside. Apart from  black holes, also other black objects, such as black strings and black rings, may live in five dimensions with near horizon geometries~$AdS_3 \times S^2$ and~$AdS_2 \times S^2 \times S^1$ respectively. Static spherically symmetric black holes in five dimensions have~$AdS_2 \times S^3$ near horizon topology and carry electric charges only~\cite{Ferrara:2006xx}. Black strings carry only magnetic charges, while black rings and stationary rotating black holes carry both electric and magnetic ones.

In order to consider all these black objects in a unified framework, we will use the black hole potential approach and the first order formalism.
As far as we know the systematic analysis of the first order formalism and the construction of the non-BPS attractor flows in five dimensions are still missing. A few attempts to study the problem were made for spherically symmetric static black holes~\cite{LopesCardoso:2007ky}. As it will be demonstrated, the result obtained by~\cite{LopesCardoso:2007ky} is a very particular case of a more general picture considered in this paper. This general analysis includes all possible charges and gauge fields (axions).

The first order formalism is important since one passes from the second order equations of motion to the first order ones, without doubling the number of them. This is related to the fact that scalar charges are not independent and the formalism automatically discards the blowing up solutions. Integrating  first order equations of motion  is certainly easier, so it enhances the possibilities to find the corresponding attractor flow. Another advantage of the first order formalism is the possibility to construct multicenter attractor flows and find walls of marginal stability~\cite{David:2009ru}. In four dimensions the relevant issue was considered in detail from many points of view~\cite{Bossard:2009we,Ceresole:2007wx,Andrianopoli:2007gt,Ceresole:2009iy,Ceresole:2009vp,Bellucci:2008sv}, but in five dimensions little attention was given to this problem. One of the purposes of the present article is to try to fill this gap.

We revisit also the 5d/4d connection. The underlying scalar manifold geometry for the~$N=2$ five dimensional supergravity is of a very special type~\cite{Gunaydin:1983bi}, which seems to be simpler than the special K\"ahler geometry of~$N=2$ supergravity~\cite{Ferrara:1997tw}. Therefore, one might na\"ively think that this fact significantly simplifies the analysis of the first order equations, eventual flat direction problems, etc. One thing to raise suspicions is the fact that the~$N=2$~$D=5$ supergravity being dimensionally reduced to~$D=4$ is equivalent to the well-known~$N=2$~$D=4$ one (see i.e~\cite{Goldstein:2007km} and refs.~therein), with the relevant special K\"ahler geometry of the scalar manifold. Therefore, the difficulties pertinent to the~$N=2$~$D=4$ supergravity are lifted up to the five dimensional theory. We will show that, in the most general setup, the five dimensional theory possesses the same level of difficulty as its four dimensional counterpart. This becomes apparent when trying to resolve attractor flow equations for BPS or non-BPS cases, as well as in finding out attractor values of the moduli or in calculating the entropy.

Nevertheless, certain simplifications do take place in special instances. Assuming the five dimensional space to be spherically symmetric, one ends up with a very simple expression for the black hole potential~\cite{Ferrara:2006xx}
$$V_5 \sim f^{ij} q_i q_j$$
with consequential simplifications of the expressions for the central charge and the fake superpotential. From the four dimensional perspective this picture corresponds to considering purely electrically charged black holes, with axions being truncated and the graviphoton charge~$p^0$ being put equal to unity.

The paper is organized as follows. In section~\ref{warmUpEx} we start with a qualitative warming up example of pure supergravities in four and five dimensions. In section~\ref{intro} we introduce the basic setup and derive the formulae we will make subsequent use of. Section~\ref{FOF} is devoted to the first order formalism in five dimensions. In section~\ref{5d4d} the 5d/4d connection is revisited from various points of view. Section~\ref{Examples} is devoted to the application of the main results to particular models and charge configurations. We end up with conclusions, where we summarize the results and give directions for future investigations.

\section{Attractors without scalars}\label{warmUpEx}
Let us start with a warming up example. The treatment of the example will be a bit cavalier, as the purpose of this section is to make a qualitative discussion and to prepare the scene. More explicit notations will be then introduced in the following sections.

We compare four and five dimensional pure supergravity extremal static black hole solutions in a spherically symmetric, asymptotically flat background. Pure supergravity models, apart from the gravitational multiplet, contain one vector field as well
\be\label{pureSugra}
{\cal S} = \int d^D x \sqrt{-g}  \left[ \frac12 \, R - \frac14\, F_{\mu\nu} F^{\mu\nu} \right].
\ee
In four dimensions the black hole under consideration is an extremal Reissner-Nordstr\"om black hole
\be\label{RN1}
ds^2 = - \left(1+\frac Mr \right)^{-2} dt^2 + \left(1+\frac Mr \right)^{2} \left( dr^2 + r^2 d\Omega_2^2\right)
\ee
with~$AdS_2 \times S^2$ near horizon geometry. Its mass and entropy are equal to
$$
M = q  , \quad S \sim  q^2
$$
where~$q$ is the electric charge of the black hole. The black hole can be dyonic, i.e. it can carry also a magnetic charge. In contrast, the five dimensional static and spherically symmetric black hole (the so-called Tangherlini black hole~\cite{Tangherlini:1963bw})
\be\label{Tangh}
ds^2 = - \left(1+\frac M{r^2} \right)^{-2} dt^2 + \left(1+\frac M{r^2} \right) \left( dr^2 + r^2 d\Omega_3^2\right)
\ee
carries an electric charge only and has~$AdS_2 \times S^3$ near horizon geometry. Its mass and entropy are given by
$$
M = q ,\qquad S \sim q^{3/2}.
$$
At first glance these black holes represent the same object in various dimensions with the following relation between the mass and the entropy:
$$
S \sim M^{\frac{D-2}{D-3}}.
$$
However, looking at them from the point of view of the underlying invariants 
one can notice that the Tangherlini black hole possesses an attractor nature unlike the Reissner-Nordstr\"om one.

The Reissner-Nordstr\"om black hole is a particular case of an
axion-dilaton family, where only the graviphoton charges are switched on and
the holomorphic prepotential is a constant. So, the underlying
invariant is the quadratic one~$I_2$ and the entropy is
$$S \sim I_2. $$
As it is widely known (see for
example~\cite{Andrianopoli:2006ub}) the axion-dilaton family, and
hence the Reissner-Nordstr\"om solution, has no five dimensional
uplift.

On the other hand, the Tangherlini black hole is a particular case of the so-called very special geometries with the underlying invariant~$I_3$. As we show below, this solution has no regular analog in four dimensions. To this end we perform the usual Kaluza-Klein reduction, singling out explicitly a non trivial~$S^1$ fibration and representing the metric~(\ref{Tangh}) in the following form:
\be\label{Tangh2}
ds^2 = - \left(1+\frac {M_5}{r} \right)^{-2} dt^2 + \left(1+\frac {M_5}{r} \right) \frac1r \left( dr^2 + r^2 d\Omega_2^2
    + r^2 \left( d\psi + \cos\theta d \phi\right)^2 \right).
\ee
Then the corresponding four dimensional metric reads
\be\label{Tangh4d}
ds^2  = - r^{1/2} \left( 1 + \frac {M_5}{r} \right)^{-3/2} dt^2 + r^{-1/2} \left( 1 + \frac {M_5}{r} \right)^{3/2} \left( dr^2 + r^2 d\Omega_2^2\right).
\ee
Upon dimensional reduction there appears a scalar field (dilaton)
\be\label{dilatEx}
e^{-2\dilaton} \sim r^{1/2} \left( 1 + \frac {M_5}{r} \right)^{1/2}.
\ee
As one can see from the expressions~(\ref{Tangh4d}) and~(\ref{dilatEx}) the corresponding four dimensional ADM mass and the dilaton diverge at infinity.

In order to be able to find regular solutions as well, one should slightly modify the Tangherlini metric~(\ref{Tangh2})
\be\label{Tangh3}
\ba{l}
\ds ds^2 = - \left(1+\frac {M_5}{r} \right)^{-2} dt^2  \\[0.5em]
\ds \phantom{ds^2 = } +\left(1+\frac {M_5}{r} \right) \left[ \left(h^0 + \frac{p^0}r \right)\left( dr^2 + r^2 d\Omega_2^2\right)
    + \frac1{h^0 + \frac{p^0}r} \left( d\psi + \cos\theta d \phi\right)^2 \right],
\ea
\ee
where~$p^0$ is a Taub-NUT charge and~$h^0$ is the distance between the center of the black hole and the core of the Taub-NUT space. The Tangherlini black hole corresponds to~$h^0=0$ and~$p^0=1$. If~$h^0$ is different from zero (hence it can be put to one without loss of generality), the spatial part of the above metric has asymptotically~${\mathbb R}^3 \times S^1$ geometry which allows us to perform a non singular dimensional reduction. The near horizon geometry of the metric~(\ref{Tangh3}) is of~$AdS_2 \times S^3/Z_{p^0}$ type and, when~$p^0=1$, it coincides with that of the Tangherlini space.

For the Tangherlini-like metric~(\ref{Tangh3}), the corresponding four dimensional picture is
\be\label{solEx}
\ba{l}
\ds ds^2 = - \left(1 +\frac{p^0}r \right)^{-1/2} \left( 1 + \frac{M_5}{r} \right)^{-3/2} dt^2
    + \left(1 +\frac{p^0}r \right)^{1/2} \left( 1 + \frac{M_5}{r} \right)^{3/2} \left( dr^2 + r^2 d\Omega_2^2\right),\\
\ds e^{-2\dilaton} \sim \frac{1+\frac{M_5}{r}}{ 1 + \frac{p^0}r}, \qquad
\ds M_4 = \frac14 \left( p^0 + 3 M_5\right)
\ea
\ee
The entropy reads
$$
S \sim \sqrt{p^0 M_5^3} .
$$
This model is the so-called~$t^3$ model, the entropy for which is a square root of the quartic invariant~$I_4 = p^0 I_3$.
For all Tangherlini-like black holes with~$p^0=1$ the entropy is proportional to the square root of the cubic invariant~$I_3$.

Despite all this, the Reissner-Nordstr\"om~(\ref{RN1}) and
Tangherlini-like metric~(\ref{Tangh2}) have one common point.
The electrically charged Reissner-Nordstr\"om metric can be obtained by
dimensional reduction from the regular Tangherlini-like one
with~$p^0 = M_5$. In this case, the metric~(\ref{solEx}) becomes of
Reissner-Nordstr\"om type and the dilaton becomes constant.

In what follows we are going to generalize the five dimensional Tangherlini-like solution including into considerations scalar fields and all possible charges. In this setup we revisit as well relations between four and five dimensional objects.

\section{Five dimensional action and black object potential}\label{intro}
Let us consider a gravity theory in five dimensions minimally coupled to~$n_s$ real scalar~$\scal^a$ and~$n_v$ vector~$A^i_M$ fields. The corresponding action has form~\cite{Ferrara:2006xx}
\be\label{action}
\ba{l}
\ds {\cal S} = \int \diff^5 x \sqrt{-g}\left[ \frac12\,R  - \frac12\, h_{ab}(\scal)\, \partial_M \scal^a \partial^M \scal^b
    - \frac14\, f_{ij}(\scal) F^i_{MN}F^{j\,MN}
\right.\\ \ds \phantom{{\cal S}=\int \diff^5 x \sqrt{g}\left[\frac12\,R\right]}\left.\hfill
    - \frac1{24}\, {\dtensor}_{ijk} \epsilon^{MNKLP} F^i_{MN}F^j_{KL} A^k_P\right]
\ea
\ee
where~the field strength is defined as~$F^i_{MN} = \der_M A^i_N - \der_N A^i_M$, the Chern-Simons contribution comes with a constant tensor~${\dtensor}_{ijk}$ to provide the gauge covariance of the action and~$M,N,K,\ldots=0,1,\ldots,4$,~$a,b = 1,\ldots,n_s$,~$i,j,k = 1,\ldots,n_v$.

The solutions we are interested in are stationary extremal black objects. Therefore, the corresponding Ansatz for the five dimensional space-time metric is chosen as
\be\label{d5metric}
\ba{l}
\ds ds^2 = g_{MN} d x^M d x^N  = e^{2\dilaton(\tau)} ds^2_{(4)}
     +  e^{2\omega (\tau)} \left(d\x - A_{\mu}^0 dx^{\mu}\right)^2,\\ [0.5em]
\ds ds^2_{(4)} = g_{\mu\nu}^{(4)} d x^\mu d x^\nu =  - e^{2U(\tau)} dt^2 + e^{-2U(\tau)}\tau^{-4}
    \left(  d\tau^2+\tau^2  \left(d\theta^2 + \sin^2 \theta d\varphi^2 \right)\right).
\ea
\ee
In these coordinates the spatial infinity corresponds to~$\tau=0$. Note that near the horizon the five dimensional metric is of the form~$AdS_2\times S^2 \times S^1$ that, generally speaking,  admits considering not only black holes but also black rings and black strings~\cite{Larsen:2006xm,Goldstein:2007km}. The black hole/string/ring choice corresponds to different fiberings with respect to the~$S^1$: if the~$S^1$ stems from fibering~$AdS_3 \sim AdS_2 \times S^1$ then the corresponding near horizon five dimensional geometry is of~$AdS_3 \times S^2$ type and, hence, describes black strings. If it instead stems from fibering~$S^3 \sim S^2 \times S^1$ then the corresponding five dimensional geometry is of~$AdS_2 \times S^3$ type and, hence, describes static black holes. Otherwise, one has either rotating black holes or black rings.

The Kaluza-Klein vector potential in the metric~(\ref{d5metric}) is chosen as
$$A^0_\mu dx^\mu = e^0_5 dt - p^0 \cos\theta\, d\varphi $$
with~$p^0$ playing the role of the Taub-NUT charge.
Due to the Einstein equations, the warp factors satisfy the following equation:
$$
\frac{d^2}{d\tau^2}\left(\omega+2\dilaton\right) + \left(\frac d{d\tau}\left(\omega+2\dilaton\right)\right)^2
    - \frac1{\tau} \frac d{d\tau}\left(\omega+2\dilaton\right) = 0
$$
that can be easily integrated
$$
e^{\omega + 2\dilaton} = \mbox{C}_1 + \mbox{C}_2\, \tau^2.
$$
For extremal black holes the integration constant~$\mbox{C}_2$ vanishes and the other constant can be normalized as~$\mbox{C}_1=1$.
In what follows we assume exactly these values for the integration constants, so that~$\omega = - 2\dilaton$.

The five dimensional vector potentials are expressed in terms of the electric field~$e^i$ and dipole charges~$p^i$ as follows:
\be\label{A5}
A^i_M dx^M = e^i_5 dt - \hat p^i \cos\theta d\varphi +a^i  d\x =  e^i dt - p^i \cos\theta d\varphi + a^i \left( d \x - A^0_\mu dx^\mu \right)
\ee
so that
\be\label{hatp}
e^i_5 = e^i - a^i e^0_5, \qquad \hat p^i = p^i - a^i p^0.
\ee
Let us note the ``lengthening'' of the dipole charges~$p^i$. As it was stressed in~\cite{Goldstein:2007km,Gao:2008hw} precisely such charges~$\hat p^i$ are genuine fluxes through the~$S^2$ and, as one can see, the axions~$a^i$ and the Taub-NUT charge contribute to them as well.
Substituting the warp factor~$\omega$ in terms of~$\dilaton$ allows us to integrate some of the Maxwell and Einstein equations
\be\label{electricPotentials}
\dot e^i_5 = - \dot a^i\,  e^0_5 + e^{2\dilaton+2U} f^{ij} \hat{q}_j, \qquad \dot e^0_5 = 2 e^{6\dilaton+2U} \hat J,
\ee
where the dots stand for a differentiation with respect to~$\tau$ and the functions~$\hat J$ and~$\hat q_i$ are defined as
\be\label{hatq}
\ba{l}
\ds\hat{q}_i = q_i - {\dtensor}_{ijk} a^j p^k + \frac12\, {\dtensor}_{ijk} a^j a^k p^0,\qquad q_i = \mbox{const},\\[0.5em]
\ds\hat J = J + a^i q_i - \frac12\, {\dtensor}_{ijk} a^i a^j p^k + \frac16\, {\dtensor}_{ijk} a^i a^j a^k p^0,\qquad J = \mbox{const}.
\ea
\ee
The constant parameter~$J$ is just an angular momentum and the~$q_i$ define electric fluxes.

Analogous expressions showed up in~\cite{Cardoso:2007rg,Goldstein:2007km,Gao:2008hw}. Here these expressions appear automatically as solutions to five dimensional Einstein and Maxwell equations and with a correct dependence on the axions and the Taub-NUT charges.

From the four dimensional perspective it is worth passing to the redefined dilaton~$\dilaton$, coupling matrices~$f_{ij}$ and~$h_{ab}$ and scalars~$\scal^a$ as follows:
\be\label{normaliz}
e^{-2\dilaton} \to 2^{1/3} e^{-2\dilaton}, \qquad f_{ij} \to 2^{-1/3} f_{ij}, \qquad \scal^a \to 2^{-1/3} \scal^a, \qquad h_{ab} \to 2^{-1/3} h_{ab}.
\ee
In terms of the redefined fields the equations of motion acquire the following form:
\be\label{eom5}
\ba{l}
\ds \ddot U = e^{2U}V_{5} ,\qquad
\ds \ddot\dilaton = \frac16\,e^{2U} \frac{\partial V_5}{\partial \dilaton}
    + \frac16\, e^{4\dilaton}\, f_{ij}\, \dot a^i  \dot a^j, \qquad
\ds \frac d{d\tau} \left[\rule{0pt}{1em} e^{4\dilaton}f_{ij}\, \dot a^j \right]  = 2 e^{2U} \frac{\partial V_5}{\partial a^i},\\[0.8em]
\ds \frac d{d\tau} \left[\rule{0pt}{1.1em} h_{ab} \dot \scal^b \right]
    = 2 \, e^{2U} \frac{\partial V_5}{\partial \scal^a}
    + \frac12\,e^{4\dilaton} \frac{\partial f_{ij}}{\partial \scal^a}\, \dot a^i  \dot a^j
    + \frac12\, \frac{\partial h_{bc}}{\partial \scal^a}\, \dot \scal^b  \dot \scal^{\dtensor} ,
\ea
\ee
while the ``energy'' constraint becomes
\be\label{firstOrderConstr}
e^{2U} V_5 = \dot U^2 + 3 \dot\dilaton^2 + \frac14\, h_{ab} \dot \scal^a \dot \scal^b + \frac14\, e^{4\dilaton} f_{ij} \dot a^i \dot a^j.
\ee
The equations of motion~(\ref{eom5}) can be derived from a one dimensional action
\be\label{effAction}
{\cal S}=\int d\tau \left[ \dot U^2 + 3 \dot\dilaton^2  + \frac14\, h_{ab}\dot \scal^a \dot \scal^b
        + \frac14\, e^{4\dilaton} f_{ij}\, \dot a^i \dot a^j + e^{2U} V_5   \right].
\ee
The five dimensional black object potential used above in eqs.~(\ref{eom5})-
(\ref{effAction}) is defined as
\be\label{V5}
V_5 = \frac12 \, e^{-6\dilaton} (p^0)^2
    + \frac12\, e^{-2\dilaton} f_{ij} \hat{p}^i \hat{p}^j
    + \frac12 \, e^{2\dilaton} f^{ij} \hat{q}_i \hat{q}_j
    + \frac12\, e^{6\dilaton} \hat J^2.
\ee

This is the genuine expression for the black object potential in
five dimensions, it contains both electric and magnetic
contributions, as well as those corresponding to the Taub-NUT charge
and rotation. Let us note that Eq.(\ref{V5}) is an expression of the black object potential previously obtained in~\cite{Goldstein:2007km,Ceresole:2007rq} in different
ways. However, quite often, rather than the full form of the potential, just some its terms (corresponding to the electric or magnetic
contributions, for example) are considered~\cite{LopesCardoso:2007ky}. In section~\ref{axTrunc} we
demonstrate that such an approach is true in some specific cases,
namely, when it is possible to neglect the axions. There it is shown
as well, how the well known~\cite{Ferrara:2006xx} five dimensional
electric potential $V_e \sim f^{ij} q_i q_j $ appears.

The black object potential
~(\ref{V5}) 
can be written in the form~\cite{Ceresole:2009id}
\be\label{V5Z}
V_5 = \frac12 Z_m^{0\,2}  + \frac12 Z^i_m f_{ij} Z^j_m + \frac12 Z_{e i} f^{ij} Z_{e j} + \frac12 Z_{e0}^2  
\ee
in terms of real electric~$Z_{e\,0}, Z_{e\,i}$ and magnetic~$Z^0_m, Z^i_m$ ``central'' charges
\be\label{realCC}
Z_m^0 = e^{-3\dilaton}  p^0, \qquad Z^i_m = e^{-\dilaton} \hat p^i, \qquad Z_{e\,i} = e^{\dilaton} \hat q_i, \qquad  Z_{e0} = e^{3\dilaton}  \hat J.
\ee

The standard four dimensional complex central charge and its covariant derivatives are then expressed in terms of the real ones~(\ref{realCC}) as follows:
\be\label{CC}
Z = \frac1{2\sqrt2} \left[ Z_0 - i \scalC^i Z_i \right] , \quad
    D_i Z = \frac1{2\sqrt2}\, e^{2\dilaton}\,\left[ -\frac{3i}2 \scalC_i \bar Z_0 + \left(\delta_i^j - \frac32 \scalC_i \scalC^j \right) \bar Z_j\right],
\ee
where~$X^i$,~$X_i$ are constrained scalar fields, to be defined below, and~$Z_0 = Z_{e0} + i Z_m^0$,~$Z_i = Z_{e\, i} + i f_{ij} Z_m^j $. In terms of these charges the black object potential can be written as
$$
V_5 = \frac12\left[ Z_0 \bar Z_0 + f^{ij} Z_i \bar Z_j\right]. 
$$

At the end of this section, let us present also the basic properties
of the very special geometry which will be intensively used in what
follows. While~$N=2$~$D=4$ supergravity is based on the special K\"ahler
geometry, the five dimensional~$N=2$ supergravity is based on the very
special geometry~\cite{Gunaydin:1983bi}. The very special geometry
occurs in M-theory compactification on Calabi-Yau sixfolds. The
supersymmetry, firstly, requires that the number~$n_s$ of the
scalars~$\scal^a$ and the number~$n_v$ of the vectors~$A^i_M$ be
related as~$n_v = n_s+1$ and, secondly it relates the coupling matrices of~$h_{ab}$ and~$f_{ij}$ in a way presented below.

All the geometry is then encoded in a cubic polynomial defined by the~$\dtensor$-tensor
\be\label{cubic}
V = \frac16\, {\dtensor}_{ijk}  \scalC^i \scalC^j \scalC^k = 1.
\ee
From the Calabi-Yau compactification point of view, $\dtensor_{ijk}$ are the intersection numbers and~$n_v$ is a Hodge number~$h_{1,1}$ (see~\cite{Cadavid:1995bk} and ref.[4] therein).
The physical scalar fields~$\scal^a$ are just a solution~$\scalC^i = \scalC^i(\scal)$ to the cubic constraint~(\ref{cubic}). The five dimensional gauge coupling matrix $f_{ij} (X)$ is given by
\begin{eqnarray}
f_{ij} (X) =  -\,\frac{\der^{\,2}  \log V }{\der \scalC^i\,\der \scalC^j}\rule[-0.75em]{0.5pt}{2em}_{\,V = 1 
} \;,
\end{eqnarray}
and hence,
\be\label{f}
f_{ij} (X) =  -  {\dtensor}_{ijk} \scalC^k + 9 \scalC_i \scalC_j , \qquad  \scalC_i = \frac16\, {\dtensor}_{ijk}  \scalC^j \scalC^k \;.
\ee
The following very special geometry identities will be useful:
\be\label{VSGident}
\ba{ll}
\ds \scalC^i \frac{\der \scalC_i}{\der \scal^a} = 0,    & \qquad
\ds h^{ab} \frac{\der \scalC^i}{\der \scal^a} \frac{\der \scalC^j}{\der \scal^b} = f^{ij} - \frac13\,\scalC^i \scalC^j, \\
\ds f_{ij} \scalC^i   = 3\scalC_j\,, & \qquad
\ds f_{ij}\frac{\der\scalC^j}{\der\scal^a} = -3\, \frac{\der \scalC_i}{\der \scal^a}.
\ea
\ee
For symmetric spaces one can introduce the inverse {\dtensor}-tensor~\cite{Gunaydin:1984ak} defined as
\be\label{cInvDef}
{\dtensor}_{i(jk}{\dtensor}_{mn)p} {\dtensor}^{ipq} = \frac43 \, \delta^q_{(j} {\dtensor}_{kmn)}.
\ee
This allows us to invert the last equation in~(\ref{f}) and rewrite the cubic polynomial~(\ref{cubic}) in terms of~$\scalC_i$
\be\label{cInv}
\scalC^i = \frac92\, {\dtensor}^{ijk} \scalC_j \scalC_k\,, \qquad {\dtensor}^{ijk} \scalC_i \scalC_j \scalC_k  = \frac29\,.
\ee
In terms of the constrained scalar fields~$\scalC^i$, the five dimensional~$N=2$ supersymmetric action~(\ref{action}) acquires the following form:
\be\label{action2}
\ba{l}
\ds {\cal S}=\int \diff^5 x \sqrt{-g}\left[ \frac12\,R - \frac12\, f_{ij}(\scalC)\, \partial_M \scalC^i \partial^M \scalC^j
    - \frac14\, f_{ij}(\scalC) F^i_{MN}F^{j\,MN}
\right.\\ \ds \phantom{{\cal S}=\int \diff^5 x \sqrt{g}\left[\frac12\,R\right]}\left.\hfill
    - \frac1{24}\, {\dtensor}_{ijk} \epsilon^{MNKLP} F^i_{MN}F^j_{KL} A^k_P\right]
\ea
\ee
so that the five dimensional~$N=2$ supersymmetric action is completely characterized by the coupling matrix~$f_{ij}(\scalC)$ and the Chern-Simons constant tensor~${\dtensor}_{ijk}$.

\section{The first order formalism}\label{FOF}
It is widely known that in four dimensions the second order equations of motion can be rewritten as first order equations. Unlike the well known analogous procedure of switching from the Lagrange to the Hamiltonian formulation of mechanics, here the number of equations does not double but remains the same~\cite{Ceresole:2007wx,Andrianopoli:2007gt}. This is related to the fact that the charges of the scalar fields might acquire specific values, in order to give a solution that is regular everywhere, from the infinity to the horizon.

Here, we would like to develop the first order formalism for black objects in five dimensions. We will concentrate mostly on the non-BPS black hole solution, since the discussion of the BPS ones is covered in the literature in much more detail.

The first order equations in the case under consideration have the following form:
\be\label{AF5}
\ba{ll}
\ds \dot U = - e^U W, & \qquad
\ds \dot \dilaton = - \frac13\,e^U \frac{\der W}{\der \dilaton},\\
\ds \dot \scal^a = - 4 \, e^U h^{ab}\frac{\der W}{\der \scal^b},& \qquad 
\ds \dot a^i = - 4 \, e^{U-4\dilaton} f^{ij} \frac{\der W}{\der a^j}.
\ea
\ee
where a real function~$W(\dilaton,X,a)$  is related to the black object potential~(\ref{V5}) in the following way:
\be\label{V5W}
V_5 = W^2
    + \frac13\, \left( \frac{\der W}{\der \dilaton}\right)^2
    + 4 h^{ab} \frac{\der W}{\der \scal^a} \frac{\der W}{\der \scal^b}
    + 4 e^{-4\dilaton} f^{ij} \frac{\der W}{\der a^i}\frac{\der W}{\der a^j}.
\ee
Let us stress that this is a genuine expression for the black object potential in terms of the superpotential~$W$ (be it fake for non-BPS or normal for BPS black holes) and it contains the full dependence on the charges, dilaton and axions. In other words, the superpotential~$W$ must satisfy this formula, in order to reproduce the correct attractor flow~(\ref{AF5}).

In the case when the Taub-NUT charge~$p^0$ vanishes one can construct an expression for the (fake) superpotential~$W$, which turns out to be equal to
\be\label{WnoTN}
W=\frac1{2\sqrt2} \left[ \pm e^{3\dilaton} \hat J + 3 e^{-\dilaton} \, p^i \scalC_i (\scal)\right].
\ee
It is not difficult to check that substituting~(\ref{WnoTN}) into~(\ref{V5W}) one gets the correct expression for the black object potential~(\ref{V5})
when~$p^0=0$. It is valid for any~$\dtensor$-geometry, be it symmetric or not. For example, it is valid for non symmetric ones~$t^3+s^3$,~$stu+t^3$ etc.
The plus sign in~(\ref{WnoTN}) corresponds to the BPS attractor flow, while the minus sign~-- to the non-BPS one.


Let us discuss further the BPS case. One can easily see, from the expressions~(\ref{CC}) and~(\ref{WnoTN}), that
$$
|Z| = \sqrt{W^2 + \frac18\, (\hat q_i \scalC^i)^2 }
$$
so formally~$W$ and~$|Z|$ are different. The regular flows corresponding to the BPS ``fake superpotential''~(\ref{WnoTN}) have the form
\be\label{BPSfake}
\ba{ll}
\ds e^{3(\dilaton - U)} = \frac{\sqrt 2}3 \dtensor_{ijk} H^i H^j H^k, & \ds \quad H_i = \dtensor_{ijk} a^j H^k,\\
\ds \scalC^i = \frac{H^i}{\left[ \frac16 \, \dtensor_{ijk} H^i H^j H^k \right]^{1/3}}, & \quad\ds  H^i \hat q_i = 0,
\ea
\ee
where~$H^i = h^i + p^i \tau$ and~$H_i = h_i + q_i \tau$.
From the last expression in the formulae above, one can easily obtain the so-called integrability condition
\be\label{integr1}
h^i q_i = h_i p^i.
\ee

When~$ \scalC^i \hat q_i$ vanishes, the superpotential~$W$ and~$|Z|$ coincide and the solution~(\ref{BPSfake}) satisfies as
well the flow equations governed by~$|Z|$. The difference between
the flows governed by~$W$ and~$|Z|$ is that the latter allows for
a much wider class of attractor flows. It is
known~\cite{Gimon:2007mh} that, for the flows governed by~$|Z|$,
the integrability condition has the form
$$h^\Lambda q_\Lambda = h_\Lambda p^\Lambda, \qquad \Lambda = 0,1,\ldots,n_v. $$
This condition coincides with~(\ref{integr1}) when~$h^0=0$.
This means that not only does~$p^0$ vanish, but the whole harmonic function~$H^0 = h^0 + p^0 \tau$ vanishes as well.

So, in the BPS case the ``fake superpotential''~(\ref{WnoTN})
describes only a particular family of the flows ($h^0=0$). Then
from~(\ref{integr1}) one ends up with a codimension one subspace
of the $2n_v$-dimensional scalar manifold, whose points determine
the asymptotic value of~$|Z|$. This situation is analogous to
having the axion truncated~$a^i=0$, which yields a~$n_v$
dimensional moduli subspace. The main difference is that here one
of the axions is expressed in terms of the other fields. The
axion-truncated case is more restrictive and may be obtained from
our case putting the~$n_v-1$  additional conditions~$h_i = 0$
(which is possible only when~$q_i=0$).

As for the non-BPS case, the expression~(\ref{WnoTN}) also describes a specific attractor flow. This form of~$W$ is not unique and in fact in section~\ref{stu} we show that there exist other forms of the fake superpotential~$W$ that depend on the values of the scalar fields at the infinity. The fact that~$W$ is not unique for the non-BPS case, noted in~\cite{Ceresole:2007wx}, implies that there might be different expressions leading to the same form of the black object potential.

In the case of vanishing~$p^0$ for the both~-- BPS and non-BPS~-- cases the entropy acquires the form
$$
 S = 2 \sqrt{| \hat J_h \dtensor_{ijk} p^i p^j p^k|},
$$
where~$\hat J_h$ is a value of~$\hat J$ in which the axions are taken on the horizon. These on-horizon values of the axions are deduced from the equation
$$
q_i = \dtensor_{ijk} a^j p^k.
$$

As we stressed, the expression~(\ref{V5W}) is a fundamental one, nevertheless quite often other relations between the black object potential~$V_5$ and superpotential~$W$ are used~\cite{LopesCardoso:2007ky}. This is related to a fact mentioned in section~\ref{intro}, when one considers not the complete form, but just particular contributions in the black object potential. Let us illustrate it on a specific example.

Let us consider the so-called electric charge configuration (that is~$q_0=p^i=0$) with vanishing axions. Then the black object potential~(\ref{V5}) acquires the following form:
\be\label{V5el}
V_5(\dilaton,\scal) = \frac12\, e^{-6\dilaton} (p^0)^2 + \frac12\, e^{2\dilaton} f^{ij}(\scal)\, q_i q_j.
\ee
On-horizon values of the dilaton~$\dilaton_h$ and the scalar fields~$\scal^a_h$  are obtained~\cite{Ferrara:1997tw} by extremizing the black object potential
$$\frac{\der V_5}{\der\dilaton} = \frac{\der V_5}{\der\scal^a} = 0 \qquad \Rightarrow \qquad
e^{-8\dilaton_h} = \frac{f^{ij}(\scal_h)\,q_i q_j} {3(p^0)^2} ,\quad \frac{\der f^{ij}}{\der \scal^a}\,q_i q_j = 0.$$
The value of the entropy is then just the critical value of the black object potential~\cite{Ferrara:1997tw}
$$S = V_5(\dilaton_h,\scal_h) =  2\,\sqrt{|p^0|}\,\left(\frac13\,f^{ij}(\scal_h) q_i q_j \right)^{3/4} $$
which coincides perfectly with~\cite{Ferrara:1996dd}.
It is clear that, in order to reproduce the correct on-horizon values of the scalars~$\scal^a_h$ one could have started with a potential of the form
\be\label{fictiousVel}
\alter{V}(\scal) \sim f^{ij}(\scal) q_i q_j
\ee
that would give the required values of the scalar fields and, naturally, would not give any information about the value of the dilaton~$\dilaton$. The correct value of the entropy would not be~$\alter{V}_{hor}$ anymore, but its  exponential~$S\sim\alter{V}_{hor}^{3/4}$. Unlike the full black object potential~(\ref{V5el}), its reduced version~(\ref{fictiousVel}) does not encode the complete behaviour of the scalar fields; it is good only for reproducing their horizon values.

Analogous considerations are valid in the first order formalism as well. Using the basic relation~(\ref{V5W}) with the axions~$a^i$ discarded, one can check that in the electric configuration the superpotential has the following form\footnote{here the plus sign corresponds to the non-BPS attractor flow, the minus sign~-- to the BPS one.}:
\be\label{Wel1}
W(\dilaton,\scal^a) = \frac1{2\sqrt2}\,\left[ \pm e^{-3\dilaton} p^0 + e^{\dilaton} q_i \scalC^i (\scal) \right]
\ee
and the on-horizon values of the scalar fields~$\scal_h^a$ and of the dilaton~$\dilaton_h$ are determined by its critical points
$$\frac{\der W}{\der\dilaton} = \frac{\der W}{\der\scal^a} = 0 \qquad \Rightarrow \qquad
e^{-4\dilaton_h} = \frac{q_i \scalC^i(\scal_h)} {3p^0} ,\quad q_i\,\frac{\der \scalC^i}{\der \scal^a} = 0.$$
The value of the entropy in the first order formalism is given by
$$S = W^2(\dilaton_h,\scal_h) = 2\, \sqrt{\strut|p^0|} \left(\frac13\,q_i \scalC^i(\scal_h) \right)^{3/2}.$$
It is obvious that the correct values of the scalar fields could have been reproduced from a ``reduced'' superpotential of the form
\be\label{fictiuosWel}
\alter{W}(\scal) \sim q_i \scalC^i (\scal).
\ee
This is exactly the expression found in~\cite{LopesCardoso:2007ky}. Between the reduced potentials~(\ref{fictiousVel}) and~(\ref{fictiuosWel}) there is a relation
\be\label{fictiousV5W}
\alter{V} = \alter{W}^2 + 3 h^{ab} \frac{\der\alter{W}}{\der\scal^a}\frac{\der\alter{W}}{\der\scal^b}
\ee
which is valid for all values of the scalars~$\scal^a$. As one sees from the equations of motion~(\ref{eom5}), the significance of the reduced potentials is quite limited: eliminating the dilaton~$\dilaton$ makes the theory valid on the horizon only. Curiously enough, the relation~(\ref{fictiousV5W}) between the reduced potentials is of the form~(\ref{V5W}), although numerical coefficients are different.

When making the above considerations we left aside many questions concerning the validity of discarding the axions~$a^i$, the dynamics of the scalar fields~$\scal^a$, the dilaton~$\dilaton$ and the warp factor~$U$ etc. We answer these questions in section~\ref{axTrunc}.

\section{5D/4D connection}\label{5d4d}
The gravitational part of the five dimensional action~(\ref{action}) contains the warp factors~$\dilaton$ and~$U$, as well as the Kaluza-Klein vector potential~$A_\mu^0$. The matter part contains~$n_s$ physical scalar fields~$\scal^a$ and~$n_v$ vector fields~$A_\mu^i$. The Kaluza-Klein vector is a four dimensional one, while the five dimensional vector fields~(\ref{A5}) are related to their four dimensional counterpart as follows:
$$A_\mu^{(4)i} d x^\mu = e^i dt - p^i \cos\theta d\varphi.  $$

The parametrization of the five dimensional metric~(\ref{d5metric}) is such that the field~$U$ is just a four dimensional warp factor, while its counterpart~$\dilaton$ defines the four dimensional dilatons in a way described below.

As it was mentioned at the end of section~\ref{intro}, in order to endow the action~(\ref{action}) with~$N=2$~supersymmetry, it is easier to write it down in terms of~$n_s+1=n_v$ scalar fields~$\scalC^i$ satisfying a cubic constraint~(\ref{cubic}), rather than in terms of the unconstrained scalars~$\scal^a$. In terms of these constrained scalars the four dimensional complex moduli~$z^i$ of~$N=2$~$D=4$ supergravity are expressed as follows:
\be\label{z}
z^i = a^i - i Y^i, \quad \mbox{where} \quad Y^i = e^{-2\dilaton} \scalC^i
\ee
so that~$a^i$ and~$Y^i$ are genuine four dimensional axions and dilatons, correspondingly.
The dilaton field~$\dilaton$ is proportional to a K\"ahler potential~$K$
$$
e^{6\dilaton}  = 8\, e^{K} = \frac{3!}{{\dtensor}_{ijk}Y^i Y^j Y^k}.
$$
The metric of the scalar manifold of the special K\"ahler geometry in~$D=4$ is then related to its~$D=5$ counterparts as
\be\label{g4}
g_{ij} = \frac{\der^{\,2}K}{\der z^i \der \bar z^j} = \frac14\,e^{4\dilaton}\,f_{ij}.
\ee
From the four dimensional perspective the five dimensional angular momentum acquires the meaning of the graviphoton electric charge
$$J = q_0 \qquad  \Rightarrow \qquad \hat J = \hat q_0.$$
With this identification the black object potential can be written as
$$V_5  = 4 e^K \left[ \hat q_0^2 + \frac14\, g^{ij} \hat q_i \hat q_j + \frac1{16}\, e^{-2K} g_{ij} \hat p^i \hat p^j + \frac1{64}\, e^{-2K} (p^0)^2 \right] $$
that coincides with the well known expression for the four dimensional black hole potential~\cite{Goldstein:2007km,Ceresole:2007rq,Ferrara:1997tw}
$$V_5 = V_{BH}. $$

The equation~(\ref{V5W}) defining the superpotential~$W$ may be rewritten in terms of the four dimensional moduli~(\ref{z})
$$V_5 = W^2 + 4 g^{ij} \frac{\der W}{\der z^i} \frac{\der W}{\der \bar z^j}. $$

\paragraph{Masses and entropies}
In four dimensions for the maximally symmetric case corresponding to the problem (that is with the spherical symmetry of the three dimensional space), the mass of a supersymmetric black hole is given by the value of the central charge~$Z_4$ at the infinity and its entropy is defined by the central charge at the horizon
$$M_4 = |Z_4|_{\infty}, \qquad  S_4 = |Z_4|^2_{hor}. $$
This means that the mass and entropy are homogeneous functions, respectively of the first and the second degree in the charges
$$M_4 \sim Q, \qquad S_4 \sim Q^2, $$
where~$Q$ is a collective notation for both electric and magnetic charges.

In five dimensions, with the assumption of the maximal symmetry (which in this case is a spherical symmetry in four dimensional space), the entropy and mass for a supersymmetric black hole are given by
$$M_5 = |Z_5|_{\infty}, \qquad S_5 = |Z_5|^{3/2}_{hor}. $$
The central charge~$Z_5$ is of the form~$q_i X^i$ or~$p^i X_i$, so it is linear in the charges~$Q$, so that the mass scales linearly, as it is in four dimensions. One might think that unlike in four dimensions the entropy in five dimensions scales as~$S_5 \sim Q^{3/2}$. The delicate point here is the maximal symmetry of the space dictated by the symmetry of the problem.

In four dimensions all electromagnetic charges~$Q_\Lambda = \{ Q_0, Q_i \}$ are black hole charges, while in five dimensions only~$Q_i$ are black hole charges and~$Q_0$ is a topological one. In the Ansatz~(\ref{d5metric}) one of the topological charges is the Taub-NUT one~$p^0$. Namely this charge is responsible for having the above mentioned maximal symmetry of the four dimensional space. When~$p^0=1$, then the four dimensional space becomes spherically symmetric; when~$p^0\neq 1$, then the corresponding four dimensional space is just ``locally'' spherically symmetric and possesses a global defect. This is reflected in the fact that one of the spherical angles does not change from~$0$ to~$4\pi$, as it should be for a spherically symmetric space, but from~$0$ to~$4\pi/p^0$.

Restoring the dependence of the entropy on the charge~$p^0$
$$S_5 =   |Z_5|^{3/2}_{hor} \sqrt{p^0}$$
one gets that the black hole entropy is again a homogeneous function of the second degree in the charges.

\subsection*{Duality group}
Let us dwell on the 4D/5D correspondence from the point of view of the duality groups~$G_4$ and~$G_5$ in four and five dimensions~\cite{Andrianopoli:2004im,Andrianopoli:2002mf,Ceresole:2009jc,Ferrara:2006yb}. We concentrate mostly on the case of~$N=8$ supersymmetry, although some results remain valid for~$N=2$, as well. For the~$N=8$ case the duality symmetry groups~$G_5$ and~$G_4$ are~$E_{6(6)}$ and~$E_{7(7)}$ correspondingly.

The Lie algebra $g_4$ of the duality group~$G_4$ can be decomposed as
follows \cite{Andrianopoli:2002mf}
$$
g_4=g_5+SO(1,1)+T_{27}(-2)+T'_{27}(+2),
$$
so that \be\label{splitting} G_4 \supset \left\{ G_5 \times
SO(1,1) \right\} \, \circledS \, T'_{27}. \ee Note that the
subgroup $SO(1,1) \circledS \, T'_{n}$ is a symmetry for any cubic
$\dtensor$-geometry (as given in (\ref{PQtransf}), (\ref{PQdil})
below). Due to this splitting, the underlying invariants of the
duality symmetry groups are related. In the case of~$G_4$ the
invariant is a quartic one~$I_4$, while for~$G_5$ it is a cubic
one~$I_3$.  The cubic invariant can be constructed in a two-fold
way
$$I_3(p) = \frac16\, d_{ijk} p^i p^j p^k, \qquad I_3(q) = \frac16\, d^{ijk} q_i q_j q_k,$$
where~${\dtensor}_{ijk}$ is exactly the tensor defining the Chern-Simons contribution in the action~(\ref{action}) and is an invariant tensor of the duality group~$G_5$. For the~$N=8$ case and for specific~$N=2$ cases the tensor~$\dtensor_{ijk}$ is such that~$\dtensor^{ijk}$ is well defined, so that~$I_3(q)$ exists.
Then, the quartic invariant is related to its cubic analog as follows:
\be\label{I4def}
I_4 = -\left(p^0 q_0 + p^i q_i \right)^2 + 4 \left[ q_0 I_3(p) - p^0 I_3(q) + \frac{\der I_3(p)}{\der p^i} \frac{\der I_3(q)}{\der q_i}\right].
\ee
Under the splitting~(\ref{splitting}) the fundamental representation of~$E_{7(7)}$ decomposes as follows:
$$ \bf 56 = 27_{+1} + 27'_{-1} + 1_{+3} + 1'_{-3}, $$
where~$\bf 27$ stands for real representations of~$E_{6(6)}$ with~$SO(1,1)$~weights~$\pm 1$ and~$\bf 1$ stands for the Kaluza-Klein singlet arising due to the dimensional reduction and having the weights~$\pm 3$. The prime denotes the contravariant representation.

On the representation~$\bf 56$ the infinitesimal transformations of~$T'_{27}$ are realised as~\cite{Andrianopoli:2004im,Andrianopoli:2002mf}
\be\label{PQtransf}
\delta \left( \ba{c} p^i \\ p^0 \\ q_i \\ q_0 \ea \right)  =  \left(
\ba{cccc}
0 &  t'^i & 0 & 0 \\
0 &  0   & 0 & 0 \\
-{\dtensor}_{ijk} t'^k  &  0 & 0 & 0 \\
0 &  0 & -t'^i & 0
\ea
\right)
\left( \ba{c} p^i \\ p^0 \\ q_i \\ q_0 \ea \right).
\ee
Another infinitesimal transformation realised on the representation~$\bf 56$ is generated by~$SO(1,1)$
\be\label{PQdil}
\delta \left( \ba{c} p^i \\ p^0 \\ q_i \\ q_0 \ea \right) =  \left(
\ba{cccc}
-\lambda &  0 & 0 & 0 \\
0 &  -3\lambda   & 0 & 0 \\
0 &  0 & \lambda & 0 \\
0 &  0 & 0 & 3\lambda
\ea
\right)
\left( \ba{c} p^i \\ p^0 \\ q_i \\ q_0 \ea \right) .
\ee
Previously in formulas~(\ref{hatq}) and~(\ref{hatp}) we defined the ``long'' charges~$\hat q_i, \hat p^i, \hat q_0$
\be\label{hatpq}
\ba{l}
\left( \ba{c} \hat p^i \\ \hat p^0 \\ \hat q_i \\ \hat q_0 \ea \right)
 =
\left(
 \ba{cccc}
  1 &   -a^i & 0 & 0\\
0 & 1 & 0 & 0\\
  - \dtensor_{ijk}a^k &   \frac12\,\dtensor_{ijk}a^j a^k & 1 & 0 \\
  - \frac12 \dtensor_{ijk} a^i a^j &   \frac16\, \dtensor_{ijk}a^i a^j a^k & a^i & 1
\ea \right)
\left( \ba{c} p^i \\ p^0 \\ q_i \\ q_0 \ea \right)
\ea
\ee
and for further simplicity we introduced here~$\hat p^0$ as well. This is just a finite form of the translational symmetry corresponding to the real representation~$\bf 27$ of the~$E_{6(6)}$, which can be easily obtained by exponentiating the nilpotent matrix of degree~3 in eq.~(\ref{PQtransf}) and putting~$t'^i = a^i$. The analogous finite transformation corresponding to~(\ref{PQdil}) with~$\lambda = \dilaton$ has the form
\be\label{PQdilFin}
\left( \ba{c} p^i \\ p^0 \\ q_i \\ q_0 \ea \right)^\prime  =  \left(
\ba{cccc}
e^{-\dilaton} &  0 & 0 & 0 \\
0 &  e^{-3\dilaton}   & 0 & 0 \\
0 &  0 & e^{\dilaton} & 0 \\
0 &  0 & 0 & e^{3\dilaton}
\ea
\right)
\left( \ba{c} p^i \\ p^0 \\ q_i \\ q_0 \ea \right) .
\ee
The simultaneous action of both the transformations~(\ref{hatpq}) and~(\ref{PQdilFin}) gives the real ``central'' charges~(\ref{realCC}), in terms of which the black object potential is expressed~(\ref{V5Z}). One may easily observe that all the dependence of these ``central'' charges on the axions is absorbed in the definition of ``long'' charges
$$Z_m^0 (p,q,\scal,a) = Z_m^0 (\hat p, \hat q,\scal,0), \quad Z_m^i (p,q,\scal,a) = Z_m^i (\hat p, \hat q,\scal,0) \quad \mbox{etc.}$$
Therefore, the same property holds for the black object potential
$$V_5 (p,q,\scal,a) = V_5 (\hat p, \hat q,\scal,0). $$
Even more, all the invariants~\cite{Ceresole:2009iy} possess the analogous property
\be\label{invHat}
i_a (p,q,\scal,a) = i_a (\hat p, \hat q, \scal, 0).
\ee
As for the quartic invariant, it remains unchanged under the transformations~(\ref{hatpq})
\be\label{I4prop}
I_4(\hat p, \hat q)
= I_4(p,q).
\ee
Even more, since~$I_4$ is a G-invariant with respect to~$E_{7(7)}$ transformations, it does not change under the simultaneous action of the transformations~(\ref{hatpq}) and~(\ref{PQdilFin})
\be\label{I4Z}\ba{l}
\ds I_4(p,q) = I_4(Z_m,Z_e) = -(Z_m^0 Z_{e0} + Z_m^i Z_{ei} )^2 + \frac23\, {\dtensor}_{ijk} Z_{e0} Z_m^i Z_m^j Z_m^k  \\
\ds \phantom{I_4(p,q) = I_4(Z_m,Z_e) = }- \frac23 \, {\dtensor}^{ijk} Z_m^0 Z_{ei} Z_{ej} Z_{ek}
    + {\dtensor}^{ijk} {\dtensor}_{ipq} Z_{ei} Z_{ej}  Z_m^p Z_m^q.
\ea
\ee
At the end of the paragraph, let us notice the following relations:
$$
\hat q_i = \frac{\der \hat q_0}{\der a^i} , \qquad
{\dtensor}_{ijk} \hat p^k = - \frac{\der \hat q_i}{\der a^j}  = - \frac{\der^2 \hat q_0}{\der a^i \der a^j}, \qquad
\frac{\der \hat p^i}{\der a^j}  = - \delta_i^j  p^0
$$
so that, in fact, there is only one essential function~$\hat q_0$, from which all the other field dependent charges can be derived.

\subsection*{$N=2$~$stu$ model and~$N=8$ supersymmetry}

The black object potential~(\ref{V5}) might possess~$N=8$ supersymmetry, depending on the form of the coupling matrix~$f_{ij}$. For the~$N=2$ case the latter is given in terms of the real special geometry, while for the~$N=8$ case this matrix is given in terms of the~$E_{6(6)}$ coset representation. However, when one considers a case with two moduli~$\scal^a$, the matrix~$f_{ij}$ becomes an~$N=2$ one (see for example~\cite{Ceresole:2009id}).

From this perspective, let us consider the~$stu$ model as a truncation of~$N=8$ supersymmetry to~$N=2$~\cite{Ferrara:2006em}. The eigenvalues of the central charge matrix are
$$
\eig_1 = Z, \qquad \eig_2 = \overline{D_{\fl s} Z}, \qquad \eig_3 = \overline{D_{\fl t} Z}, \qquad \eig_4 = \overline{D_{\fl u} Z},
$$
where the overline denotes the complex conjugation and the hat stands for the flat indices
$$D_{\fl i} Z = \tetrad_{\fl i}^i D_i Z $$
with respect to the einbeins~$\tetrad_{\fl i}^i$
$$g^{i\bar i} = \tetrad^i_{\fl i} \eta^{\fl i \bar{\fl i}} \tetrad^{\bar i}_{\bar{\fl i}}
    =  \mbox{diag} \left[-(s-\bar s)^2,-(t-\bar t)^2,-(u-\bar u)^2 \right]$$
of  the K\"ahler metric of the~$stu$ model. Using the relations~(\ref{g4}) and~(\ref{cfstu}) we choose the einbeins to be\footnote{another choice of phases is possible, as well.}
$$\tetrad^i_{\fl i} = 2 i  e^{-2\dilaton} \, \mbox{diag} \left[  \scalC^1 ,  \scalC^2,  \scalC^3 \right], \qquad
\tetrad^{\bar i}_{\bar{\fl i}} = \overline{\tetrad^i_{\fl i}}.$$
Then one can easily show~\cite{Ferrara:2006em} that the attractor equations
$$\der_i V_4 = 2 \bar Z D_i Z + i C_{ijk} g^{i \bar i} g^{j \bar j} \overline{D_j Z}\, \overline{D_k Z}, \qquad
C_{ijk} = e^K \dtensor_{ijk}$$
acquire the form
$$\bar\eig_1 \bar\eig_2 + \eig_3 \eig_4 = 0, \quad
\bar\eig_1 \bar\eig_3 + \eig_2 \eig_4 = 0, \quad
\bar\eig_1 \bar\eig_4 + \eig_2 \eig_4 = 0.$$
There are two types of solutions to these equations:
\begin{enumerate}
\item just one of~$\eig_i$ is different from zero. If this is to be~$\eig_1$, then this solution corresponds to a BPS one, otherwise this solution is non-BPS with vanishing central charge.
\item all~$\eig_i$ are different from zero, so that this solution is a non-BPS one. In this case the absolute values of all~$\eig_i$ are equal, and the sum of their phases is equal to~$\pi$.
\end{enumerate}

The above considerations refer to the attractor point, but something can be said about~$\eig_i$ in the whole space. For this, let us consider an electric configuration without axions. Then
$$
Z = \frac i{2\sqrt2} \left[ Z_m^0 - \scalC^i Z_{e\,i} \right], \qquad
D_i Z = \frac{e^{2\dilaton}}{2\sqrt2} \left[ -\frac32 \scalC_i Z_m^0 + \left(\delta_i^j - \frac32 \scalC^j \scalC_i \right) Z_{e\,j}\right].
$$
Having a very special geometry origin, the flat indices of~$Z_{e\,i}$ are defined by using the metric~$f_{ij}$, which in our case has the form~(\ref{cfstu}). Therefore,
$$Z_{e\,\fl i} = \scalC^i  Z_{e\,i} \quad \mbox{(no summation over~$i$)}. $$
This means that the above introduced eigenvalues can be written in the following form (in the whole space, not just at the horizon!)~\cite{Ferrara:1997ci}:
$$\ba{ll}
\ds \eig_1 = \frac i{2\sqrt2} \left[ Z_m^0 - Z_{e\,\fl 1} - Z_{e\,\fl 3} - Z_{e\,\fl 3}\right], &
\ds\qquad \eig_2 = \frac i{2\sqrt2} \left[Z_m^0 - Z_{e\,\fl1} + Z_{e\,\fl2} + Z_{e\,\fl3} \right],\\
\ds\eig_3 = \frac i{2\sqrt2} \left[Z_m^0 + Z_{e\,\fl1} - Z_{e\,\fl2} + Z_{e\,\fl3} \right], &
\ds\qquad \eig_4 = \frac i{2\sqrt2} \left[Z_m^0 + Z_{e\,\fl1} + Z_{e\,\fl2} - Z_{e\,\fl3} \right].
\ea $$

\section{Examples}\label{Examples}
Here we specify the results obtained in the previous sections for particular cases. We concentrate on symmetric~$\dtensor$-geometries. Particularly, we investigate models which, after dimensional reduction, yield the so-called~$t^3$ and~$stu$ models. Some well known charge configurations are revisited in more detail as well.

\subsection{$t^3$ model}
In this section we consider a five dimensional pure supergravity. As it was mentioned in section~\ref{warmUpEx} it generalizes the Tangherlini black hole and, after dimensional reduction, yields the well-known~$t^3$ model in four dimensions.
In this case the action~(\ref{action}) contains no scalar fields and a single vector field. The coupling matrices are given by
\be\label{cft3}
{\dtensor}_{111} = 6, \qquad {\dtensor}^{111} = \frac29\,, \qquad f_{11} = 3
\ee
and the single modulus of the~$t^3$ model is expressed through the warp factor~$\dilaton$ and the fourth component~$a^1$ of the vector potential~$A^1$ in eq.~(\ref{A5}) as follows:
\be\label{t3mod}
t = a^1 - i\,e^{-2\dilaton}.
\ee
One can check that the five dimensional black object potential~(\ref{V5}), after plugging eqs.~(\ref{cft3}) and~(\ref{t3mod}), yields exactly the black hole potential of the~$t^3$ model.

We are mostly interested in the non-BPS branch, so the fake superpotential in this case is given by~\cite{Ceresole:2009iy}
$$
W = \frac{(-I_4)^{1/4}}{4\sqrt{\Mf}} \left[\rule{0pt}{1.2em} \Mf^2 + 3\, \Nf^2 + 3 \right]
$$
where~$\Mf$ and~$\Nf$, in terms of the five dimensional dilaton and gauge field, are defined as follows:
$$
\ba{l}
\ds \Mf = \frac12\, \frac{e^{2\dilaton} \left( \nu  \hat{\sigma}^+ + \hat{\sigma}^-\right)^2 + e^{-2\dilaton} \left( \nu - 1\right)^2 }
    {\nu \left(\hat{\sigma}^+ + \hat{\sigma}^- \right)}, \\
\ds \Nf = \frac12\, \frac{e^{2\dilaton} \left( \left(\nu  \hat{\sigma}^+\right)^2 - \left(\hat{\sigma}^-\right)^2\right) + e^{-2\dilaton} \left( \nu^2 - 1\right) }
    {\nu \left(\hat{\sigma}^+ + \hat{\sigma}^- \right)},
\ea
$$
with~$\hat\sigma^\pm$ and~$\nu$ defined by thes formula
\be\label{sigma_nu_t3}
\ds {\hat\sigma}^\pm = \frac12\, \frac{\sqrt{\strut -I_4} \pm (\ds  p^0 \hat q_0 + \frac13\, \hat p^1 \hat q_1)}{\ds({\hat p}^1)^2 - \frac13\, p^0 \hat q_1}, \quad
\nu^3 = \frac{2\,{(\hat p^1)}^3 +  p^0 \left(\ds\sqrt{\strut-I_4} -  p^0 \hat q_0 - \frac13\, \hat p^1 \hat q_1 \right)}
    {2\,{(\hat p^1)}^3 -  p^0 \left(\ds \sqrt{\strut -I_4} +  p^0 \hat q_0 + \frac13\, \hat p^1 \hat q_1\right)}
\ee
and, finally, the so-called quartic invariant~$I_4$~(\ref{I4def}) is equal to
$$I_4 = - (p^0 q_0 + p^1 q_1)^2 + 4\,  (p^1)^3 q_0 - \frac4{27}\, p^0 q_1^3 + \frac43 \, (p^1 q_1)^2 . $$

The flow of the scalar fields is given by
\be\label{t3flow}
e^{-4U}  = \h_0 \h_1^3 - c^2 , \qquad
e^{-2\dilaton} = 2\,\nu \,\frac{ e^{-2U}}{\znam} \,\frac{\sqrt{\strut-I_4}}{(p^1)^2 - \frac13\, p^0 q_1}, \qquad
a = \frac{\chisl}{\znam},
\ee
where we introduced the harmonic functions~$\h_{\Lambda}(\tau) = {\dtensor}_\Lambda + (-I_4)^{1/4}\tau$ and defined~$S,R$ as follows:
$$
\ba{l}
\chisl =  \h_1^2 (\sigma^+ \nu - \sigma^-) (\nu + 1) + \h_0 \h_1 (\sigma^+ \nu + \sigma^-) (\nu - 1)
    + 2c (\sigma^+ \nu^2 + \sigma^-),\\[0.5em]
\znam = \h_1^2 ( \nu + 1)^2  + \h_0 \h_1 (\nu - 1)^2 + 2 c (\nu^2 - 1).
\ea
$$
Let us mention here that this solution can be obtained from that found in~\cite{Cvetic:1996xz}.

At the horizon the scalar fields acquire the following values:
$$e^{-2U_h}  = \sqrt{\strut -I_4} \tau^2 + O(\tau), \qquad
    e^{-2\dilaton_h} = \frac{\nu}{\nu^2 + 1} \, \frac{\sqrt{\strut -I_4}}{(p^1)^2 - \frac13 \, p^0 q_1}, \qquad
    a_h = \frac{\sigma^+ \nu^2  - \sigma^-}{\nu^2+1}.
$$
It is easy to see that the Tangherlini solution~(\ref{Tangh2}) is a particular case of the general solution~(\ref{t3flow}) with~$p^1=q_0=0$,~$p^0=1$ and~$c=d_0=0$. This choice of the constants corresponds to the electrical configuration with the unit Taub-NUT charge~$p^0$ and vanishing integration constants~$c$ and~$d_0$.

The horizon values of the functions~$\Mf$ and~$\Nf$ are
$$
\Mf=1, \quad \Nf=0,
$$
so the entropy is
$$
S_5= W_{H}=\sqrt{\strut -I_4}.
$$
When the magnetic charge~$p^1$ vanishes, one can easily obtain that the formula for the entropy acquires the form
$$
S_5 = \sqrt{\strut (p^0)^2 J^2-4p^0I_3(q)}.
$$
This formula is analogous to the one given by~\cite{Gaiotto:2005xt,Gaiotto:2005gf,Hornlund:2010tr,Borsten:2009zy}. Let us mention that~$p^0$ can be either positive or negative.

In the case of vanishing Taub-NUT charge~$p^0$ the scalars behave as follows:
$$e^{-2\dilaton}  = \frac12\, \sqrt{\frac{\h_0}{\h_1} - \left( \frac{c}{\h_1^2} \right)^2} \, \frac{\sqrt{\strut-I_4}}{(p^1)^2}, \qquad
a = \frac{q_1}{6p^1} + \frac{c}{2\h_1^2}\, \frac{\sqrt{\strut-I_4}}{(p^1)^2}
$$
and the expression for the fake superpotential simplifies significantly 
$$
W = \frac1{2\sqrt{2}} \left( 3 \, e^{-\dilaton}  p^1 -  e^{3\dilaton} \hat q_0 \right).
$$
This is exactly the expression obtained when setting~$s=t=u$ and making the corresponding charge identifications in the expression~(\ref{Wstu}). In addition, the expressions~(\ref{sigma_nu_t3}) are obtained in analogous way from the more general expressions~(\ref{sigmanu}).

\subsection{$stu$ model}\label{stu}
The four dimensional~$stu$ model corresponds to the case when the five dimensional action~(\ref{action}) contains two scalar fields~$\scal^a$ and three vector fields~$A_M^i=\{ A_M^1, A_M^2,A_M^3\}$. The couplings are defined as follows:
\be\label{cfstu}
{\dtensor}_{123}=1, \quad {\dtensor}^{123}=1, \quad f_{ij} =\left(
\ba{ccc}
\ds (X^1)^{-2} & 0 & 0 \\
0 & \ds(X^2)^{-2} & 0 \\
0 & 0 & \ds (X^3)^{-2}
\ea\right),
\ee
where the newly introduced scalar fields~$X^i$ satisfy the cubic constraint~(\ref{cubic})
$$
\scalC^1 \scalC^2 \scalC^3 = 1 \quad \Rightarrow \quad \scalC^i = \scalC^i (\scal^1,\scal^2).
$$
The moduli~$z^i$ of the~$stu$ model are then given by
$$
z^i =  a^i - i \,e^{-2\dilaton} \scalC^i.
$$

The fake superpotential for the~$stu$ model is obtained in~\cite{Ceresole:2009vp,Bellucci:2008sv} and, in terms of the five dimensional fields, it can be represented in the following way:
\be\label{Wstu}
\ba{l}
\ds W = \frac{(-I_4)^{1/4}}{4\sqrt{\Mf_1 \Mf_2 \Mf_3}}\left[\rule[-0.3em]{0pt}{1.6em} \Mf_1 \Mf_2 \Mf_3 + \Mf_1 \Nf_2 \Nf_3 + \Nf_1 \Mf_2 \Nf_3
    + \Nf_1 \Nf_2 \Mf_3
\right.\\
\ds \phantom{W = \frac{(-I_4)^{-1/4}}{4\sqrt{\Mf_1 \Mf_2 \Mf_3}}}
\left.\rule[-0.3em]{0pt}{1.6em}
    + \Mf_1 + \Mf_2 + \Mf_3\right],\\
\ds\phantom{W=} \Mf_i = \frac12\, \frac{3\,  e^{2\dilaton}\,\scalC_i  \left( \hat\nu_i \hat \sigma^+_i + \hat\sigma^-_i \right)^2
    + e^{-2\dilaton}\, X^i \left( \hat\nu_i - 1 \right)^2 }
    {\hat \nu_i (\hat\sigma^+_i+\hat\sigma^-_i)},\\[0.75em]
\ds\phantom{W=} \Nf_i = \frac12\, \frac{3\, e^{2\dilaton}\,\scalC_i  \left( ( \hat\nu_i \hat\sigma^+_i )^2 - ( \hat\sigma^-_i )^2 \right)
    + e^{-2\dilaton}\,\scalC^i \left( \hat\nu_i^2 - 1 \right) }
    {\hat \nu_i (\hat\sigma^+_i+\hat\sigma^-_i)}.
\ea
\ee
Here~$\Mf_i$ and~$\Nf_i$ are defined in terms of~$\hat\sigma^\pm_i$ and~$\hat\nu_i$, which are functions of the gauge fields~$a^i$ only and have the following form:
\be\label{sigmanu}
\ba{ll}
\ds \hat \sigma^\pm_i = \sigma^\pm_i\; \rule[-0.9em]{0.04em}{1.7em}_{\raisebox{0.7em}{\scriptsize$\ba{l}p\to\hat p\\
    q\to\hat q\ea$}}, \quad
\ds \hat \nu_i = \nu_i\; \rule[-0.9em]{0.04em}{1.7em}_{\raisebox{0.7em}{\scriptsize$\ba{l}p\to\hat p\\
    q\to\hat q\ea$}}, &\quad
\ds \nu_i = e^{\alpha_i} \nu, \quad 
\sum_i \alpha_i = 0, \\
\ds \sigma^{\pm}_i = \frac{\sqrt{\strut-I_4} \pm \left(p^\L q_\L - 2 p^i q_i \right)}{{\dtensor}_{ijk}p^i p^j - 2 p^0 q_i},&\quad
\ds \nu^3 = \frac{2 p^1 p^2 p^3 + p^0 \left( \sqrt{\strut -I_4} - p^\L q_\L \right)}
    {2 p^1 p^2 p^3 - p^0 \left(\sqrt{\strut -I_4} + p^\L q_\L \right)}.
\ea
\ee
As it is shown in~(\ref{I4prop}) the quartic invariant remains unchanged when substituting the ``long'' charges. Quite simple properties hold for~$\sigma^\pm_i$ and~$\nu$ as well
$$
\hat\sigma^\pm_i = \sigma^\pm_i \pm a^i, \qquad \hat\nu = \nu.
$$

The attractor flows for the scalar fields~$X^i$ and the warp factor~$U$ have the form~\cite{Ceresole:2009vp,Bellucci:2008sv}
$$
\ba{l}
\ds e^{-4U}  = \h_0 \h_1 \h_2 \h_3 - c^2, \qquad \h_\Lambda = {\dtensor}_\Lambda + (-I_4)^{1/4}\tau,\\
\ds a^i = \frac{\chisl_i}{\znam_i}, \quad
\scalC^i = \frac{\nu_i}{\nu} \,
    \left[
    \frac{(\sigma^+_i + \sigma^-_i)^2 \znam_j \znam_l}{(\sigma^+_j + \sigma^-_j)(\sigma^+_k + \sigma^-_k)\znam_i^2}
    \right]^{1/3}, \quad
e^{-2\dilaton}  = 2\nu e^{-2U}\,\prod_{i=1}^3 \left[\frac{ \sigma^+_i + \sigma^-_i }{\znam_i}\right]^{1/3}
\ea
$$
where for the sake of brevity we abbreviated\footnote{we suppose that all the indices~$i,j,k$ are different and~$j<l$.}
$$
\ba{l}
\chisl_i =  \h_j \h_l (\sigma^+_i \nu_i - \sigma^-_i) (\nu_i + 1)
    + \h_0 \h_i (\sigma^+_i \nu_i + \sigma^-_i) (\nu_i - 1)
    + 2c (\sigma^+_i \nu_i^2 + \sigma^-_i),\\[0.5em]
\znam_i = \h_j \h_l ( \nu_i + 1)^2  + \h_0 \h_i (\nu_i - 1)^2 + 2 c (\nu_i^2 - 1).
\ea
$$
Along the flow the functions~$\Mf_i$ and~$\Nf_i$, in terms of which the fake superpotential~$W$ is defined, acquire the following form:
$$\Mf_i = \frac12\, {\dtensor}_{ijk} \h_j \h_k e^{2U}, \qquad \Nf_i = c\, e^{2U}. $$

At the horizon one gets the following values of the scalar fields:
$$
\ba{l}
\ds a^i = \frac{\sigma^+_i \nu_i^2 - \sigma^-_i}{\nu_i^2 + 1}, \quad
e^{-2\dilaton} = \nu\,\prod_{i=1}^3\left[\frac{\sigma^+_i + \sigma^-_i}{\nu_i^2+1}\right]^{1/3}, \\
\ds \scalC^i = \frac{\nu_i}{\nu}\,\left[\frac{(\nu_j^2 +1)(\nu_l^2 +1)}{(\sigma^+_j + \sigma^-_j)(\sigma^+_k + \sigma^-_k)}\right]^{1/3}
    \left[\frac{\sigma^+_i + \sigma^-_i}{\nu_i^2 +1}\right]^{2/3}\,.
\ea
$$
At the same time the functions~$\Mf_i$ and~$\Nf_i$, in terms of which the fake superpotential~$W$ is defined, and the entropy read
$$\Nf_i = 0, \qquad \Mf_i = 1, \qquad S_5= W_{H}=\sqrt{\strut -I_4}. $$
Once again, when the magnetic charges~$p^1,p^2,p^3$ vanish, one can easily obtain the following formula for the entropy:
$$
S=\sqrt{\strut (p^0)^2 J^2-4p^0I_3(q)}.
$$
This formula is the non-BPS analog of the well-known formula for rotating electrically charged black holes (see, e.g.~\cite{Gaiotto:2005xt,Gaiotto:2005gf,Borsten:2009zy} and reference therein). Here~$p^0$ can also be either positive or negative.

Let us mention that from this consideration one can easily obtain
analogous expressions for the~$st^2$ model and, as it was already
said, for the~$t^3$ model as well.

\subsection{Some charge configurations}\label{axTrunc}
In this section we consider charge configurations which allow us to truncate axions. From eq.~(\ref{eom5}), one sees that, in order to have a consistent axion truncation, the potential~$V_5$ should not contain linear terms in the axions
\be\label{noLinTerm}
\frac{\der V_5}{\der a^i}\rule[-0.7em]{0.4pt}{2em}_{\;a=0} = 0.
\ee
This is possible when all three conditions hold simultaneously
\be\label{threeConds}
p^i p^0 = 0, \qquad q_i q_0 = 0, \qquad p^i q_j = 0 \quad \mbox{for all}\quad i,j,
\ee
with the possible solutions corresponding to large black holes
\be\label{CCsol}
p^i = q_0 = 0 \quad (a) \qquad \mbox{or}\qquad q_i = p^0 = 0 \quad (b) \qquad \mbox{or}\qquad q_i = p^j = 0 \quad (c)
\ee
Let us consider each of these cases in detail.

\paragraph{Electric configuration} This case is the best known in five dimensions, due to the fact that five dimensional spherically symmetric and static black holes can carry electric charges only.  Spherical symmetry requires that, along with the axions being equal to zero, the Taub-NUT charge is unity. Nevertheless, we will keep~$p^0$ arbitrary for the moment. Here, we demonstrate how a potential~$W \sim q_i X^i $ obtained in~\cite{LopesCardoso:2007ky} appears.

As it was mentioned in section~\ref{FOF}, the black object potential for this case acquires the form~(\ref{V5el})
and the corresponding fake superpotential is defined by eq.~(\ref{Wel1}). The equations of motion~(\ref{eom5}) can be rewritten in the following way:
\be\label{eomEl}
\ddot v = 2 e^{2v} (p^0)^2,\qquad
\ddot u = e^{2u} V_e, \qquad
\ds \frac{d}{d\tau}\left[\strut h_{ab} \dot\scal^b \right] = \frac32\,e^{2u} \frac{\der V_e}{\der\scal^a}
    + \frac12\,\frac{\der h_{bc}}{\der \scal^a} \dot\scal^b\dot\scal^{\dtensor}
\ee
where we introduced new variables~$v=U-3\dilaton$,~$u=U+\dilaton$ and the electric potential is defined as~\cite{Ferrara:2006xx}
$$
V_e = \frac23\, f^{ij}q_i q_j.
$$
One sees that the degree of freedom corresponding to~$v$ decouples and the appropriate equation of motion can be easily integrated out
\be\label{solu}
e^{-v} = \sqrt2 \, H^0(\tau),  \qquad H^0(\tau) \equiv h^0 +  p^0 \tau, \qquad h^0 = \mbox{const},
\ee
while the rest of the equations can be derived from the action
\be\label{eqElectr}
\ds {\cal S} = \int \diff \tau \left[ \dot u^2 + \frac13\, h_{ab} \dot \scal^a \dot \scal^b + e^{2u} \, V_e\right].
\ee
The first order equations of motion of this action can be represented in the form
\be\label{eomElfirstOrder}
\dot u = - e^u W_e, \qquad \dot\scal^a = - 3 h^{ab} e^u \frac{\der W_e}{\der \scal^b}
\ee
with the superpotential~$W_e$ defined through the relation
\be\label{We}
V_e = W_e^2 + 3 h^{ab} \frac{\der W_e}{\der \scal^a}  \frac{\der W_e}{\der \scal^b}.
\ee

Using the relation~(\ref{VSGident}) one can show that the following form of the superpotential
$$W_e = \frac{\sqrt2}3\, q_i \scalC^i (\scal) $$
satisfies the definition~(\ref{We}). This is a formula found in~\cite{LopesCardoso:2007ky} for the electric configuration with axions equal to zero. 

The first order equations~(\ref{eomElfirstOrder}) can be integrated in terms of the constrained scalars~$\scalC_i(\scal)$
$$\scalC_i = \frac{\sqrt2}3\, e^u H_i(\tau), \qquad H_i(\tau) = h_i + q_i \tau.$$
For symmetric spaces, by means of the constraint~(\ref{cInv}), the previous expression may be written in terms of~$\scalC^i$
$$
\scalC^i = \left( \frac92 \right)^{1/3} \, \frac{{\dtensor}^{ijk} H_j H_k}{\left({\dtensor}^{mnl} H_m H_n H_l\right)^{2/3}}, \qquad
    e^{-3u} = \frac{\sqrt2}3 \, {\dtensor}^{ijk} H_i H_j H_k
$$
so that the metric acquires the form
\be\label{electricMetric}\ba{l}
\ds ds^2 = - \left[ \frac23\, {\dtensor}^{ijk}  H_i H_j H_k \right]^{-2/3} dt^2 \\
\ds\phantom{ds^2 =}
    + \left[ \frac23\, {\dtensor}^{ijk}  H_i H_j H_k \right]^{1/3} \left(  H^0 \tau^{-4}(d\tau^2 +\tau^2 d\Omega_2^2) + \frac1{H^0} ({d\x} + p^0 \cos\theta d\varphi)^2 \right)
\ea
\ee
The five dimensional black hole entropy can be expressed in different ways
$$ S = V_5  = W^2  = 2^{1/4} \sqrt{|p^0|}  V_e^{3/4} = 2^{1/4} \sqrt{|p^0|}  W_e^{3/2}$$
where all functions are evaluated at the horizon.
Using the Bekenstein-Hawking formula, from the metric~(\ref{electricMetric}) one can calculate the black hole entropy in terms of the charges
$$ S = \lim_{\tau \to \infty} \frac{\sqrt{\strut \frac23\, {\dtensor}^{ijk} H^0 H_i H_j H_k}}{\tau^2}  =  \sqrt{\frac23\, p^0 {\dtensor}^{ijk} q_i q_j q_k}. $$
The above expression coincides with the entropy of a four
dimensional electrically charged black hole. As in four
dimensions, it is also a homogeneous function of degree two in the
charges, but there is a difference in the physical interpretation
of the charges. In four dimensions all charges are the charges of
the black hole. Instead, in five dimensions~$p^0$ is a topological
charge characterizing the space-time. Most often this charge was
chosen to be unity~\cite{Ferrara:2006xx} to have a full spherical
symmetry and~$S^3$ geometry on the horizon. Setting~$p^0=1$ we get
the well known expression for the entropy in terms of the cubic
invariant~$I_3$
$$ S = 2 \sqrt{I_3}.$$
As for the ADM mass, according to the definition, it is proportional to the Christoffel symbol~$\Gamma^0_{10}$. For this case it turns out to be equal to~$W_{e\, \infty}$ and, in terms of the charges, it looks like
$$M_5 = {\dtensor}^{ijk} q_i h_j h_k.$$
Note that it does not depend on~$p^0$.

The corresponding four-dimensional metric~(\ref{d5metric}) has the form
$$ds_{(4)}^2 = -\frac{dt^2}{\left(\frac23\,  H^0 {\dtensor}^{ijk} H_i H_j H_k \right)^{1/2}}
    + \left(\frac23\,  H^0 {\dtensor}^{ijk} H_i H_j H_k \right)^{1/2} \tau^{-4} (d\tau^2 + \tau^2 d\Omega_2^2) $$
so that the four dimensional ADM mass is equal to
$$ M_4 = \frac{p^0}{h^0} + {\dtensor}^{ijk} q_i h_j h_k $$
and in contrast to its five dimensional counterpart, it depends on~$p^0$.

The near horizon metric
can be represented in the form
$$ds^2 = - \left(\frac23\, {\dtensor}^{ijk} q_i q_j q_k\right)^{-2/3} \frac{dt^2}{\tau^2}
     + p^0\,\left(\frac23\, {\dtensor}^{ijk} q_i q_j q_k\right)^{1/3}\left[ \frac{d\tau^2}{\tau^2} + 4\, d\Omega_{3/p^0}^2 \right], $$
which is nothing but~$AdS_2 \times S^3 / Z_{p^0}$, where the division by~$p^0$ means that the space possesses a global defect characterized by the fact that one of the angular coordinates changes from~$0$ to~$4\pi/p^0$ rather than to~$4\pi$.

To summarize, one can say that the five and four dimensional entropies coincide, even though the nature of the charges is different, while the masses are different.

\paragraph{Magnetic configuration}
Let us now consider the solution~(\ref{CCsol}b). This configuration is dual to the electric one with the duality transformations~$p\leftrightarrow q$,~$\dilaton \to - \dilaton$,~$f_{ij} \leftrightarrow f^{ij}$. The black object potential has the form
$$
V_5 = \frac12\, e^{-2\dilaton} f_{ij} p^i p^j +\frac12\,e^{6\dilaton} q_0^2
$$
and the equations of motion~(\ref{eom5}) become
$$
\ddot v = 2\, e^{2v} q_0^2 ,\qquad \ddot u = e^{2u} V_m, \qquad
    \frac{d}{d\tau}\left[ h_{ab} \dot\scal^b \right] = \frac32 e^{2u}\, \frac{\der V_m}{\der \scal^a}
    + \frac12\, \frac{\der h_{bc}}{\der\scal^a} \dot\scal^b \dot\scal^c
$$
where we introduce new variables~$u=U-\dilaton$,~$v=U+3\dilaton$ and the magnetic potential~$V_m$ is defined as
\be\label{Vmag}
V_m = \frac23\,  f_{ij} p^i p^j.
\ee
As in the electric case, here too, the degree of freedom corresponding to~$v$ decouples and the appropriate equation of motion can be easily integrated out
\be\label{solvMag}
e^{-v} = \sqrt2 H_0(\tau), \qquad H_0(\tau) = h_0 + q_0 \tau
\ee
and the only independent fields that remain are~$\scal^a$ and~$u$, whose dynamics is governed by the action
$$
\ds {\cal S} = \int \diff \tau \left[\dot u^2 + \frac13\, h_{ab} \dot \scal^a \dot \scal^b + e^{2u}V_m\right].
$$
The first order  equations of the attractor flow and the definition of the fake superpotential in this case are in fact identical to those~(\ref{eomElfirstOrder}),~(\ref{We}) for the electric configuration
\be\label{eomMagfirstOrder}
\dot u = - e^u W_m, \qquad \dot\scal^a = - 3 h^{ab} e^u \frac{\der W_m}{\der \scal^b}
\ee
upon obvious changing of the notations~$V_e\to V_m$ and~$W_e\to W_m$, with the following form of the fake superpotential:
$$W_m = \sqrt2\, p^i \scalC_i (\scal).$$
The first order equations~(\ref{eomMagfirstOrder}) can be integrated
$$ \scalC^i = \frac{H^i}{\left(\frac16\, {\dtensor}_{jkl} H^j H^k H^l \right)^{1/3}}, \qquad e^{-3u} = \frac{\sqrt2}3 \, {\dtensor}_{jkl} H^j H^k H^l $$
and the metric acquires the following form:
$$
\ba{l}
\ds ds^2 = \left[\frac13\, {\dtensor}_{jkl} H^j H^k H^l \right]^{2/3}  \tau^{-4} \left(d\tau^2 + \tau^2 d\Omega_2^2 \right) 
    + \frac{2H_0}{ \left( \frac13\, {\dtensor}_{jkl} H^j H^k H^l \right)^{1/3}}  \left[ {d\x}^2 - \frac1{H_0} {d\x} dt \right]
\ea
$$
The black hole entropy and the five dimensional ADM mass are equal to
$$S = \lim_{\tau\to\infty} \frac{\sqrt{\strut \frac23 \, {\dtensor}_{ijk} H_0 H^i H^j H^k}}{\tau^2} = 2\sqrt{\frac16\, {\dtensor}_{ijk} q_0 p^i p^j p^k}\,, \qquad
 M_5 = 2 {\dtensor}_{ijk} h^i h^j p^k 
.$$
The four dimensional part of the metric is given by
$$ds_{(4)}^2 = -\frac{dt^2}{\sqrt{\strut \frac23 \, {\dtensor}_{ijk} H_0 H^i H^j H^k}} + \sqrt{\strut \frac23 \, {\dtensor}_{ijk} H_0 H^i H^j H^k}
    \,\tau^{-4} (d\tau^2 + \tau^2 d\Omega_2^2) $$
so that the four dimensional ADM mass is equal to
$$ M_4 = \frac{q_0}{h_0} + 2 {\dtensor}_{ijk} h^i h^j p^k 
. $$
Near the horizon the metric acquires the form
$$\ba{l}
\ds ds^2 = \left[ \frac13\, {\dtensor}_{ijk} p^i p^j p^k \right]^{2/3} \left[ d\Omega_2^2
    + \frac{d\tau^2}{\tau^2}
    + \frac{2 q_0}{\frac13\, {\dtensor}_{ijk} p^i p^j p^k} \left( d\x^2 - \frac1{q_0 \tau} d\x d t\right)\right]
\ea
$$
and corresponds to~$AdS_3/Z_{q_0} \times S^2$ geometry.

As we see, in this configuration again the five and four dimensional entropies are equal, while the masses are different.

\paragraph{$D0-D6$ configuration}
In the case~(\ref{CCsol}c) the black object potential
\be\label{D0D6V5}
V_5 =\frac12\, e^{-6\dilaton} (p^0)^2 + \frac12\, e^{6\dilaton} q_0^2
\ee
ceases to depend on the scalar fields~$\scal^a$, so that they completely become decoupled from the background
\be\label{scalEOM}
\frac d{d\tau} \left[ h_{ab} \dot \scal^b \right] =  \frac12\, \frac{\partial h_{bc}}{\partial \scal^a}\, \dot \scal^b \dot \scal^c,
\ee
while the warp factor~$U$ and the dilaton~$\dilaton$ satisfy the following equations of motion:
\be\label{eomD0D6}
\ds \ddot U = \frac12\, e^{2(U-3\dilaton)} (p^0)^2 +\frac12\, e^{2(U+3\dilaton)} q_0^2,\qquad 
\ds \ddot \dilaton = - \frac12\, e^{2(U-3\dilaton)} (p^0)^2 + \frac12\,e^{2(U+3\dilaton)} q_0^2.
\ee
The latter can be obtained from the action
$$
{\cal S} = \int \diff \tau \left[\dot U^2 + 3 \dot \dilaton^2 + e^{2U} V_5 \right]
$$
with the potential~$V_5$ given by eq.(\ref{D0D6V5}). Since the scalar fields~$\scal^a$ are now decoupled from the background, they do not enter any more into the first order equations
\be\label{D0D6foe}
\dot U = - e^U W, \qquad \dot\dilaton = - \frac13\, e^U \frac{\der W}{\der\dilaton}
\ee
where the fake superpotential~$W$ is defined through the formula
$$
V_5 = W^2 + \frac13 \, \left( \frac{\der W}{\der \dilaton}\right)^2
$$
and, naturally, it does not depend on the scalar fields~$\scal^a$:
\be\label{WD0D6}
W = \frac1{2\sqrt2} \left[ q_0^{2/3} e^{2\dilaton} + (p^0)^{2/3} e^{-2\dilaton} \right]^{3/2}.
\ee
The first order equations~(\ref{D0D6foe}) can be integrated, and the unique regular solution has the following form:
$$
e^{6\dilaton} = \left|\frac{p^0}{q_0}\right| \left( \frac{H^2 - c^2}{H^2 + c^2}\right)^{3/2}, \quad
e^{-4U} = H^4 - c^2, \quad H = h + (-I_4)^{1/4}\, \tau, \quad I_4 = -(p^0 q_0)^2.
$$
In the case when the constant~$c$ vanishes, one can get rid of one of the degrees of freedom as it occurred in the electric and magnetic cases. With the degree of freedom corresponding to~$\dilaton$ being eliminated, the potential and the fake superpotential become constant
$$V_{D0D6} = |p^0 q_0|, \qquad W_{D0D6} = \sqrt{|p^0 q_0|} $$
coinciding with their minimal values.

Concerning the scalar fields, the only regular solution occurs when they get constant values~$\scal^a=const$ all along the flow. This situation is similar to that of pure supergravity, since the scalar fields are completely decoupled and acquire constant values.

To end this paragraph, let us write down 
the black hole entropy
$$S = \lim_{\tau\to \infty} e^{-2U}\tau^{-2} = |p^0 q_0| $$
which is nothing but a minimum of the potential~(\ref{D0D6V5}) or of the square of the fake superpotential~(\ref{WD0D6}).

\section{Conclusions}
In this paper we revisited five dimensional Einstein-Maxwell Chern-Simons supergravity.
We investigated black objects with all charges and gauge fields (axions) switched on. For these objects we derived the equations of motion and the corresponding potential governing the dynamics of the system. The potential, which is a sum of squares of ``central charges''~(\ref{V5Z}), was previously obtained by brute force in~\cite{Ceresole:2009id}.

We constructed and investigated the first order formalism for black objects in five space-time dimensions. A first result of the investigation is presented in the eq.~(\ref{V5W}) above, where the expression of the black object potential is given in terms of the (fake) superpotential. It was shown that our consideration encompasses all previously known cases~\cite{Lopes Cardoso:2007ky,Ceresole:2007rq}.

The relation between solutions for five and four dimensional black objects is revisited. Group and symmetry properties are investigated
. As one sees from formulae~(\ref{invHat})-(\ref{I4Z}), all
duality invariants~$i_a$ and symplectic invariant~$I_4$ and can be
rewritten in terms of either the so-called ``long'' charges or the
real central charges.

Some examples have been investigated here. Firstly, as a warming
up example 4d/5d pure supergravities were analyzed. Despite the
fact that Reissner-Nordstr\"om and Tangherlini solutions belong to
different classes, they have one point in common. Under specific
values of the parameters, the Tangherlini-like solution can be
reduced to a pure electric Reissner-Nordstr\"om one.

The other examples represent models with a very special geometry.
The case of vanishing Taub-NUT charge~$p^0$ is investigated in
detail. Eq.~(\ref{WnoTN}) above displays a second result of our investigation,
i.e. the explicit expression of the (fake) superpotential,
corresponding to the (non) BPS attractor flow. An important fact
is that in this case the fake superpotential is constructed for an
arbitrary~$\dtensor$-tensor, so that it is  valid also for non
symmetric scalar manifolds.

We have investigated special cases of
symmetric~$\dtensor$-geometry (i.e. the so-called~$t^3$ and~$stu$
models) in full generality. The non-BPS branch of these examples
was studied: the fake superpotentials, attractor flows and horizon
values were written down. This yields the third result
of this work, as shown in eq.~(\ref{Wstu}) for the~$stu$ model,
where all limiting cases, such as the~$t^3$~model or particular
charge configurations, are encompassed. In the end we showed that,
in the absence of axions, the previously known results emerge.

It is interesting to generalize the obtained results for the case of multicenter black objects and, hence, investigate the issue of the marginal stability. From this point of view it is also interesting to study the possibility of obtaining a 4d single center solution from a 5d multicenter one upon dimensional reduction.

\section*{Acknowledgments}
We would like to thank Prof. Mario Trigiante for enlightening
discussions and criticism. A. Y. would like to thank CERN PH-TH
Department for his hospitality during the final stage of the
project.
 This work is supported by the
ERC Advanced Grant no. 226455, ``\textit{Supersymmetry, Quantum
Gravity and Gauge Fields}'' (\textit{SUPERFIELDS}) and in part by
MIUR-PRIN contract 20075ATT78 and DOE Grant DE-FG03-91ER40662.

\end{document}